\documentclass[useAMS,usenatbib]{mn2e}
\pdfoutput=1

\usepackage{hyperref}
\usepackage{aas_macros}
\usepackage{tikzexternal}
\newcommand{\pgfplotsset}[1] {}
\newcommand{\flash}{{\sc FLASH}}
\newcommand{\mvect}[1]{\mathbf{#1}}
\newcommand{\ud}{\mathrm{d}}
\newcommand{\grad}[1]{\nabla #1}
\newcommand{\vsgr}{V4046 Sgr}
\newcommand{\dqtau}{DQ Tau}
\newcommand\kms{\ifmmode{\rm km\thinspace s^{-1}}\else km\thinspace s$^{-1}$\fi}
\newenvironment{DIFnomarkup}{}{}
\graphicspath{{./figures/}}
\DeclareGraphicsExtensions{.pdf,.png,.jpg}

\title[Circumbinary Gas Flows in Close T Tauri Binaries]
{Modelling Circumbinary Gas Flows in Close T Tauri Binaries%
\thanks{Based on observations collected at the European Southern Observatory.}}

\author[M. de Val-Borro, G. F. Gahm, H. C. Stempels and A. Pepli\'{n}ski]
{M.~de~Val-Borro,$^1$\thanks{E-mail:
\href{mailto:deval@mps.mpg.de}{\url{deval@mps.mpg.de}}}
G.~F.~Gahm,$^2$ H.~C.~Stempels,$^3$ and A.~Pepli\'{n}ski$^2$\\
$^{1}$Max Planck Institute for Solar System Research, DE-37191
Katlenburg-Lindau, Germany\\
$^{2}$Stockholm University, AlbaNova University Center, SE-10691,
Stockholm, Sweden\\
$^{3}$Department of Physics and Astronomy, Uppsala University, Box 515,
SE-75120 Uppsala, Sweden
}

\begin{document}

\date{Received May 10, 2010; }

\pagerange{\pageref{firstpage}--\pageref{lastpage}} \pubyear{2010}

\maketitle

\label{firstpage}

\begin{abstract}
Young close binaries open central gaps in the surrounding circumbinary
accretion disc, but the stellar components may still gain mass from gas
crossing through the gap.  It is not well understood how this process
operates and how the stellar components are affected by such inflows.
Our main goal is to investigate how gas accretion takes place and
evolves in close T Tauri binary systems. In particular, we model the
accretion flows around two close T Tauri binaries, \vsgr{} and \dqtau{},
both showing periodic changes in emission lines, although
their orbital characteristics are very different.
In order to derive the density and velocity maps of the circumbinary
material, we employ two-dimensional hydrodynamic simulations with a
locally isothermal equation of state. 
The flow patterns become quasi-stable after a few orbits in the frame
co-rotating with the system. Gas flows across the circumbinary gap
through the co-rotating Lagrangian points, and local circumstellar discs
develop around both components. Spiral density patterns develop in the
circumbinary disc that transport angular momentum efficiently.
Mass is preferentially channelled towards the primary and its
circumstellar disc is more massive than the disc around the secondary.
We also compare the derived density distribution to observed line
profile variability.
The line profile variability tracing the gas flows in the central cavity
shows clear similarities with the corresponding observed line profile
variability in \vsgr{}, but only when the local circumstellar disc
emission was excluded.  Closer to the stars normal magnetospheric
accretion may dominate while further out the dynamic accretion process
outlined here dominates.  Periodic changes in the accretion rates onto
the stars can explain the outbursts of line emission observed in
eccentric systems such as \dqtau{}.
\end{abstract}

\begin{keywords}
Accretion, accretion discs --
binaries: close --
hydrodynamics --
methods: numerical -- 
stars: individual: \vsgr{}, \dqtau{} --
stars: pre-main sequence.
\end{keywords}

\section{Introduction}

Classical T Tauri stars (CTTS) are young pre-main-sequence objects with
pronounced emission line features and an infrared excess indicative of
dust in the circumstellar discs. A number of CTTS are confirmed close
binaries, with orbital periods from days to weeks, and several of these
show emission lines that vary in intensity and line shape with phase. In
such systems, the stars orbit in a gap opened by tidal interactions
inside a circumbinary disc. Eccentric systems, like \dqtau{}
\citep{1997AJ....114..781B} and UZ Tau E \citep{2005A&A...429..939M},
show enhanced emission line activity close to periastron passages. This
indicates that the accretion in such systems is non-axisymmetric, and
perturbed by the orbital interaction with the inner disc. However, in at
least one system with equal-mass components and nearly circular orbits,
namely \vsgr{}, emitting gas fills the gap between the stars and the
surrounding disc.  Observations of \vsgr{} by
\citet{2004A&A...421.1159S} provided evidence that the flows are
manifestations of non-axisymmetric mass accretion in this system. These
observations show that gas flows across the gap also in young, close
binary systems with circular orbits.

Circumbinary discs are common in a wide range of astrophysical objects
such as young binary stellar systems and massive binary black hole
systems at the centre of galaxies \citep[see
e.g.][]{2007PASJ...59..427H}.  Binary stars are believed to form by
fragmentation of dense cores and accrete mass from the envelope via a
circumbinary disc.  It is of interest to understand how the components
in binaries evolve due to preferential mass accretion, and how the
orbital elements may change with time.  Protobinary systems allow us to
study disc evolution under well defined conditions, since the sizes of
the central gaps in the circumbinary disc are determined by tidal
truncation and the stellar components are supposed coeval.  The common
evolution of the circumstellar discs is governed by accretion from the
circumbinary disc that can extend up to hundreds of AU.

Numerical simulations of binary systems with a circumbinary disc based
on Smoothed Particle Hydrodynamics (SPH)
\citep{1994ApJ...421..651A,1997MNRAS.285...33B} and grid-based methods
\citep{2002A&A...387..550G,2004A&A...423..559G} show that an inner
cavity forms inside of the 2:1 resonance.  \citet{1997MNRAS.285...33B}
found that, in binaries with a mass ratio different from unity, the less
massive protostar accretes more material and is therefore more luminous.
However, that claim has been later refuted using high-resolution
two-dimensional grid-based simulations
\citep{2005ApJ...623..922O,2007AstL...33..594S,2010ApJ...708..485H}.
Recent SPH simulations of multiple star formation in molecular clouds
have shown that unequal mass binaries are rare
\citep{2004MNRAS.351..617D,2008ASPC..390...76C}, although observational
data indicates that mass ratios different from unity are common
\citep{1997AJ....113.2246R}.  For reviews of observational properties
and numerical simulations of young multiple systems see
\citet{2000prpl.conf..703M,2007prpl.conf..379D,2007prpl.conf..133G}.

\citet{1996ApJ...467L..77A} found that binaries with eccentric orbits
can generate non-axisymmetric gas flows from the disc edge in agreement
with the periodic line changes observed in such systems. Preliminary
simulations of systems with circular orbits (\citealt{2005.PA} -- see
\citealt{2006Ap&SS.304..149G}, page 151), have shown that mass accretion
occurs, but the gas density inside the gap is much lower than for
systems with eccentric orbits.  Such simulations, and with similar
conclusions, were later done by \citet{2007AstL...33..594S}. Systems
with companions on inclined circular orbits have also been modelled
\citep{1995MNRAS.274..987P}.

In the present paper we investigate accretion in close T Tauri binaries
and follow the early stages of their evolution.  We carry out numerical
simulations of protoplanetary discs surrounding close T Tauri binaries
using an Adaptive Mesh Refinement (AMR) scheme, providing high spatial
resolution inside the inner cavity opened by the binary orbit.  We
follow the formation of a circumstellar gap in the disc and map the
accretion flows onto the stellar components where local circumstellar
accretion discs form, and estimate line profile emission from the
accretion flows.  We select the orbital parameters of the systems to
match two well observed spectroscopic binary systems \vsgr{} (circular)
and \dqtau{} (eccentric) in order to compare the accretion process in
these two different cases, and we explore how our results compare with
earlier simulations of similar systems.  Observations indicate that
accretion from such circumbinary discs occurs over the gap but that the
gas flows are non-axisymmetric.  The stellar components in these systems
gain most of their mass by accretion through circumstellar discs.  Our
calculations also provide information on whether mass accretion is
directed preferentially to the primary or the secondary in the different
systems. In order to shed some light on this question we have also
modelled a system on a circular orbit, but with a mass ratio different
from unity.  Finally, both \vsgr{} and \dqtau{} show remarkable periodic
changes in the Balmer line profiles. We explore in a qualitative way to
what extent these variations could match predictions from our model.
This study is also relevant to understand accretion onto a planet that
is massive enough to tidally open a gap in a circumstellar disc. Giant
planets can accrete material outside the planet's orbit through the
annular cavity at a rate comparable with the accretion rate onto the
inner disc \citep{2001ApJ...547..457K}.

This paper is organized as follows.  Our physical model and numerical
code are described in Section~\ref{sec:setup}.  The main results from
our simulations of circular and eccentric binary systems are presented
in Section~\ref{sec:results}, and comparisons to observed emission line
variations in \vsgr{} and \dqtau{} are made in
Section~\ref{sec:comparison}, where we also discuss differences between
accretion in systems with different mass ratios. Finally, the
conclusions are given in Section~\ref{sec:conclusions}.

\section{Numerical setup}
\label{sec:setup}

We consider the evolution of a non-self-gravitating circumbinary disc
around close T Tauri binary systems with circular and eccentric orbits.
First, we model a circular system with parameters similar to the T Tauri
system \vsgr{}, for which we have acquired complementary observations.
The parameters stellar mass, semi-major axis, period, inclination and
eccentricity are given in Table~\ref{table:v4046}.  In the second set of
simulations we study close binary systems with moderate orbital
eccentricity and mass ratio close to unity such as \dqtau{} (see
Table~\ref{table:DQTau}).  There is observational evidence that both
system are surrounded by extensive circumbinary discs \citep[see ][for
details]{2004A&A...421.1159S,1997AJ....113.1841M,2008A&A...492..469K}.

\subsection{Basic equations}

The system is described by the two-dimensional Navier-Stokes equations
using vertically-integrated variables
\begin{eqnarray}
\frac{\partial\Sigma}{\partial t} + \nabla \cdot (\Sigma \mvect{v}) & = & 0 \\
\frac{\partial\mvect{v}}{\partial t} + (\mvect{v} \cdot \nabla ) \mvect{v} &
= & - \frac{1}{\Sigma}\grad{P} - \grad{\Phi}
\label{eq:NS}
\end{eqnarray}
$\Sigma$ denotes the surface density, $P$ is the pressure, ${\mathbf v}$
is the velocity of the fluid and $\Phi$ is the gravitational potential.
The surface density can be expressed as
\begin{equation}
    \Sigma(r,\phi) = \int_{-\infty}^\infty \rho(r,\phi,z) \, \ud z,
\end{equation}
where $\rho$ is the three-dimensional density.
The gravitational potential is given by the formula
\begin{equation}
    \phi (r,t) = \sum_{i=1}^2 \phi_i (r,t).
\end{equation}
Each stellar component has a potential of the form
\begin{equation}
\phi_i = \frac {-G M_i}{\sqrt{(r-r_i)^2 + \epsilon^2}},
\end{equation}
where $\epsilon$ is the gravitational softening, $r$ is the distance
from the centre of mass and $r_i$ denotes the position of the star. The
softening length is typically $0.1a$ in our simulations. This value is
smaller than the Hill radius where the gravity from the star dominates
and the gas forms a circumstellar disc.  Therefore, the choice of
softening length does not affect strongly the results of our simulations.

We adopt a simple equation of state for the gas in our models with a
temperature profile depending on the distance to the centre of mass and
to each stellar component.  A locally isothermal solver for the
circumbinary disc with constant aspect ratio $H/r = 0.05$ is
implemented, where $H$ is the disc scale height and $r$ is the distance
from the barycentre \citep[see e.g.,][]{1985prpl.conf..981L}.  The sound
speed in the circumbinary disk is given approximately by the formula
\begin{equation}
  c_\rmn{s}= H \Omega_\rmn{d},
\end{equation}
where $H$ is the scale-height and $\Omega_\rmn{d}$ is the Keplerian
angular velocity in the circumbinary disc.  Each stellar component has a
temperature distribution according to the prescription in
\citet{2008MNRAS.386..164P}.  Accretion processes in circumbinary discs
can be an efficient mechanism to convert gravitational potential energy
into radiation for high mass accretion rates.  In our case, the disc is
assumed to radiate efficiently the thermal energy generated from tidal
dissipation, viscous heating and stellar radiation.  In the absence of
an efficient cooling mechanism the gas would heat up and the disc would
become geometrically thick.  We do not include irradiation effects from
the stars in our equation of state.

Self-gravity of the disc was not considered in our simulations since the
total mass of the disc material in the computational domain is much
smaller than the mass of the binary system \citep[see
e.g.][]{2004A&A...423..559G}.  Initially, we kept the system in a fixed
Keplerian orbit and assumed a coplanar circumbinary disc for simplicity.
In some computations, we did not account for the accretion of material
onto the stars and therefore a high density peak forms around the stars.
Such high-density peaks are clearly an artefact of our simulations and
are not real. We therefore exclude these regions when doing a comparison
with observations (see Section~\ref{sec:comparison}).

\begin{table}
\caption{Orbital parameters of the circular binary system \vsgr{}
\citep{2004A&A...421.1159S,2000IAUS..200P..28Q}.
Masses are in units of the solar mass and distances
are in solar radii.
}
\label{table:v4046}
\centering
\begin{tabular}{c c c}
\hline
Parameter & Primary & Secondary \\
\hline
$M$  &    0.912 $M_\odot$     &   0.873 $M_\odot$\\
$a_1, a_2$  &    4.52 $R_\odot$      &  4.72 $R_\odot$\\
$P$  &   \multicolumn{2}{c}{2.4213459 d} \\
$i$  &   \multicolumn{2}{c}{$35^\circ$}  \\
$e$  &   \multicolumn{2}{c}{$\le0.01$} \\
\hline
\end{tabular}
\end{table}

\begin{table}
\caption{Orbital parameters of the elliptic binary system
\dqtau{} \citep{1997AJ....113.1841M}}.
\label{table:DQTau}
\centering
\begin{tabular}{c c c}
\hline
Parameter & Primary & Secondary \\
\hline
$M$  &    0.55 $M_\odot$     &   0.55 $M_\odot$\\
$a_1, a_2$  &    6.6 $R_\odot$      &  6.6 $R_\odot$\\
$P$  &   \multicolumn{2}{c}{15.8043 d} \\
$i$  &   \multicolumn{2}{c}{$23^\circ$}  \\
$e$  &   \multicolumn{2}{c}{$0.556$} \\
\hline
\end{tabular}
\end{table}

\subsection{Initial and boundary conditions}

\pgfplotsset{map style/.style=
    {enlargelimits=false,axis on top, width=7.5cm, height=7.5cm,
    xlabel = {$x\ [R_\odot]$}, ylabel = {$y\ [R_\odot]$},
    colorbar style={x=0.3cm} }}

\pgfplotsset{horizontal style/.style=
    {enlargelimits=false,axis on top, width=7.5cm, height=7.5cm,
    xlabel = {$x\ [R_\odot]$}, ylabel = {$y\ [R_\odot]$},
    colorbar style={y=0.3cm} }}

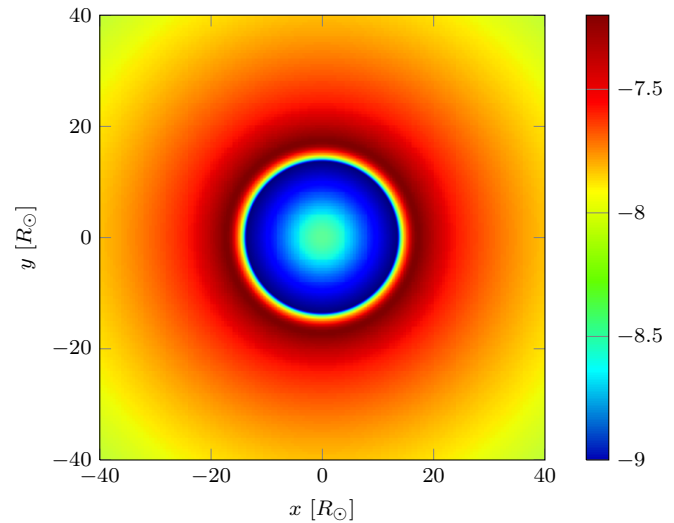
\begin{figure}
  \centering
  \begin{tikzpicture}
    \begin{axis}[colorbar right,colormap/bluered,map style]
      \addplot[mesh,samples=150,domain=-9:-7.2]
      {x};
      \addplot graphics
	[xmin=-40,xmax=40,ymin=-40,ymax=40]
   	{init_prof_pgf};
    \end{axis}
  \end{tikzpicture}
  \caption{Initial surface density distribution with coordinates
  expressed in solar radii for the \dqtau{} case (see
  Table~\ref{table:DQTau}).  The surface density in computational
  units is given in logarithmic scale. The outer circumbinary disc is
  represented by the equilibrium distribution of an accretion disc
  around a single star, where the surface density is proportional to
  $r^{-0.5}$ \citep{2002A&A...387..550G}.
  }
  \label{fig:initprof}
\end{figure}

The binary orbits and mass ratios assumed in the presented simulations
are chosen to closely match the configurations of \vsgr{} and \dqtau{}.
At time zero we start with a surface density proportional to $r^{-0.5}$
following \citet{2002A&A...387..550G}.  However, the density
distribution at large distances in the circumbinary disc has no
appreciable effect on the accretion flows through the gap.  We created a
central axisymmetric cavity around the centre of mass of the stellar
system inside the 2:1 resonance initially, where we have set the density
to $\Sigma \approx 10^{-2} \Sigma_0$ in our computational units.  The
circumbinary gas rotates in the same direction as the stars and
initially has Keplerian velocities around the centre of mass of the
system.  In Fig.~\ref{fig:initprof}, we show an example of an initial
axisymmetric surface density profile with a power-law radial dependence
at large distances.  The gravitational potential of the stars is
introduced smoothly over half an orbit to avoid the formation of strong
shocks.

We used open outflowing boundaries with wave damping zones outside $r=2a$
to avoid wave reflection and mass loss. Mass inflow into our
computational domain was not considered. The damping regions were
implemented in the outer regions of the circumbinary disc
where the following equation was solved after each time step:
\begin{equation}
\frac {\ud x}{\ud t} = - \frac {x-x_\mathrm{0}} {\tau} R(r),
\end{equation}
where $x$ represents either the surface density, or the  velocity components,
$\tau$ is the orbital period of the gas 
at $r=2a$ and $R(r)$ is a ramp-up function which becomes unity
at $r=3a$ \citep{2006MNRAS.370..529D}.
With this boundary prescription we can avoid
mass loss through the open boundaries.  Simulations were run
using different damping functions to confirm that the gap
structure and accretion streams do not depend strongly on our
choice of boundary conditions.

Material from the circumbinary disc is accreted through the gap and
forms circumstellar discs.  Accretion onto the stars is estimated by
removing material from a region inside the Hill radius for each stellar
component every time-step \citep[see
e.g.][]{2008MNRAS.386..164P,2009ApJ...700.1148D}, although the dynamical
mass of the stars is kept constant. Since the obtained accretion rates
and time scales are small this is a reasonable assumption.
The mass is removed from the circumstellar disk after each time step
using the expression
\begin{equation}
   \Delta \Sigma = \max\left(0,\Sigma -
   \Sigma_\mathrm{av} \right),
   \label{eq:acckley}
\end{equation}
where $\Sigma_\mathrm{av}$ denotes the average density in the
star's neighbourhood
$r_\mathrm{acc}<|\mathbf{r}-\mathbf{r}_\mathrm{s}|< 2r_\mathrm{acc}$,
where $r_\mathrm{acc}$ is a fraction of the softening length.
The momentum of the accreted material is removed from the system.

\subsection{Numerical code}

We performed two-dimensional hydrodynamic simulations 
of the material in the orbital plane on a Cartesian grid.
The simulations were run on a dynamically refined grid for
$\sim10$ orbital periods when the system has reached
a quasi-static configuration.

The Cartesian version of
the \flash{} code was run in the inertial frame
centered on the centre of mass of the binary system.
\flash{} is a parallel block-structured code based on the
Piecewise Parabolic Method (PPM)
\citep{2000ApJS..131..273F}
\footnote{\flash{} is available at
\href{http://www.flash.uchicago.edu/}{http://www.flash.uchicago.edu/}}.
The code has been extensively tested in various compressible
flow astrophysical problems \citep[see e.g.][and references
therein]{2006MNRAS.370..529D}.
\flash{} uses the PARAMESH library to refine the grid
dynamically based on the behaviour of the solution.

For this work, we explicitly ensure the conservative transport
of angular momentum to a high degree of accuracy introducing the Coriolis 
force as source terms \citep{1998A&A...338L..37K}.  This is particularly
important when large density gradients are present in the computational
domain.  In our problem, we need to resolve a large density contrast
between the inner cavity and the circumbinary disc.  Our implementation
includes an adapted equation of state module and a simple N-body solver
to evolve the binary orbit.
Physical viscosity is not included in our simulations.  In addition,
\flash{} is able to resolve the spiral arms in our simulations and an
artificial viscosity is not needed to smooth out the shocks.

The unit of distance is $a=1$ in the simulations, and the unit of time
is the orbital period of the system which is given by
\begin{equation}
    P = 2 \pi \sqrt{\frac{a^3}{G(M_1+M_2)}}=2 \pi,
\end{equation}
where $G (M_1+M_2)=1$ for computational convenience.  In our units the
angular frequency of the binary is $\Omega= 1$.

The number of cells in our simulations is $n_\mathrm{x} \times
n_{y}=(256,256)$ and $n_\mathrm{x} \times n_{y}=(512,512)$ for the
higher-resolution simulations, with four additional levels of
refinement.  Therefore, we achieve high-resolution inside the
circumbinary gap and in the vicinity of the stars.  The computational
domain extends between $-3a \le x \le 3a$ and $-3a \le y \le 3a$, with
uniform spacing in our base grid in both directions.

\section{Results}
\label{sec:results}

\begin{figure}
  \centering
  \begin{tikzpicture}[scale=.8]
    \tikzstyle{post}=[->,shorten >=1pt,>=stealth',thick]
    \foreach \x in {+,-}
    {
      \fill[color=lightgray] (\x3,\x-4) .. controls (\x2,0) ..
      (\x1,\x0.15) -- (\x1,\x0.25)..  controls (\x2,0) ..  (\x3,-\x1) -- cycle;
    }
    \shade[inner color=darkgray, even odd rule] (0,0) circle (5.) (0,0) circle (3.);
    \fill[color=gray] (1,0) circle (.3);
    \fill[color=gray] (-1,0) circle (.3);
    \draw[color=gray,very thick] (-1,.3) .. controls (-.4,.3) and (.4,-.3) .. (1,-.3);
    \draw (0,0) circle (1);
    \draw[post] (1,0) arc (0:60:1);
    \draw[post] (-1,0) arc (180:240:1);
    \draw[post] (-15:3.5) arc (-15:15:3.5);
    \draw[post] (165:3.5) arc (165:195:3.5);
    \draw plot[mark=x] (-1.75,0) node [left] {$L_2$};
    \draw plot[mark=x] (1.75,0) node [right] {$L_3$};
    \draw[post] (1.75,0) arc (0:60:1.75);
    \draw[post] (-1.75,0) arc (180:240:1.75);
    \shade[ball color=white] (1,0) circle (.15);
    \shade[ball color=black] (-1,0) circle (.15);
    \draw (0,3.5) node {Circumbinary disc};
    \draw (1,-.5) node {Primary};
    \draw (-1,.5) node {Secondary};
  \end{tikzpicture}
  \caption{ Diagram of a gravitationally bound binary system on a
  circular orbit surrounded by a circumbinary disc in Keplerian
  rotation. The gravitational interaction creates a central cavity at
  the tidal truncation radius.  Local circumstellar discs form around
  each star from material accreted from the circumbinary disc through
  the Lagrangian points L$_2$ and L$_3$.  }
  \label{fig:circumbinarydisc}
\end{figure}
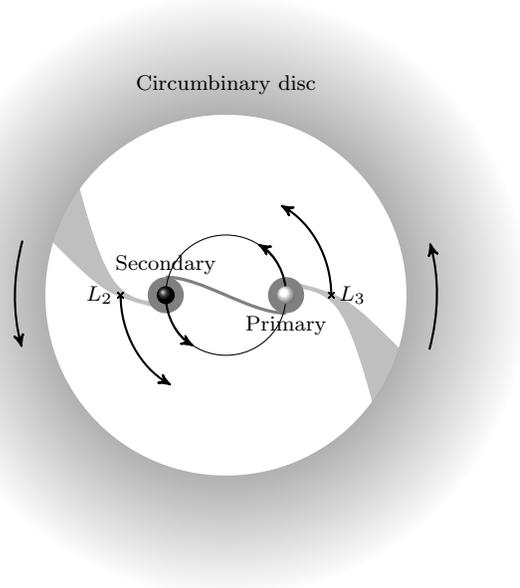

In this section we discuss the results of our simulations of circular
and eccentric binary systems surrounded by a circumbinary disc.  A
central cavity forms on the dynamical scale by the gravitational effect
of the binary with the inner edge close to the 2:1 corotation resonance.
The disc relaxes viscously on short time scales and the flow becomes
quasi-stationary in the corotating frame for circular binaries.  This
case is also appropriate to test the stability of our numerical method.
The accretion flows form global patterns as shown in
Fig.~\ref{fig:circumbinarydisc} for a binary on a circular orbit with
a surrounding circumbinary disc after the envelope has been dissipated.
Gas streams from the surrounding disc across the central cavity through
the co-linear Lagrangian points L$_2$ and L$_3$, and small local discs
develop around each star.  

\subsection{Circular Binaries}
\label{sec:circular}

\begin{figure}
  \centering
  \begin{tikzpicture}
    \begin{axis}[colorbar right,colormap/bluered,map style]
      \addplot[mesh,samples=150,domain=-15:-8]
      {x};
      \addplot graphics
	[xmin=-20,xmax=20,ymin=-20,ymax=20]
   	{binary_2d_5lev_hdf5_chk_0076_pgf};
    \end{axis}
  \end{tikzpicture}
  \caption{Surface density map in logarithmic scale for
  a simulation of the system \vsgr{} after 5
  orbits including accretion onto the stars. 
  The initial surface density of the circumbinary
  disc is unity.
  The secondary is located at $(x,y)=(-4.72,0)\ R_\odot$, and
  the primary at $(x,y)=(4.52,0)\ R_\odot$.
  The system rotates in counterclockwise direction.
  }
  \label{fig:dens}
\end{figure}
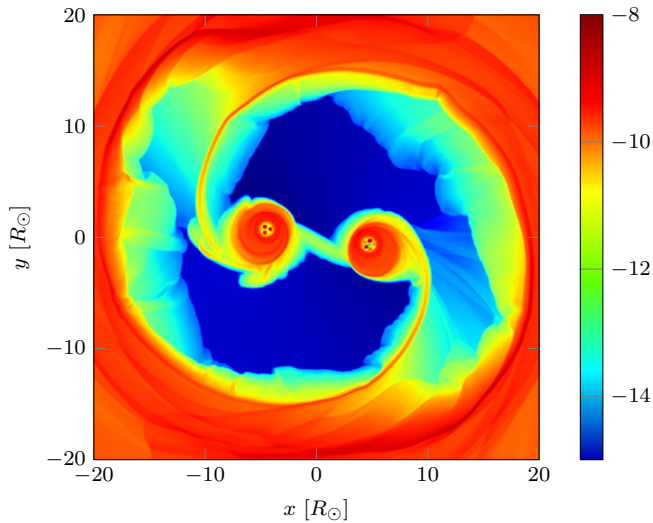

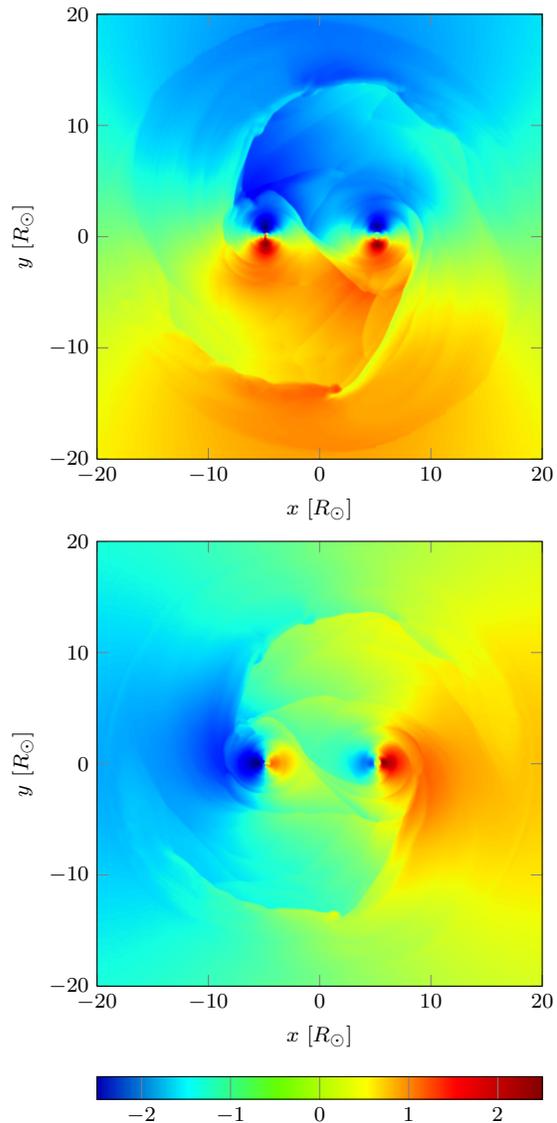
\begin{figure}
  \centering
  \begin{tikzpicture}
    \matrix {
    \begin{axis}[map style]
      \addplot graphics
	[xmin=-20,xmax=20,ymin=-20,ymax=20]
   	{velx0500_pgf};
    \end{axis} \\
    \begin{axis}[colorbar horizontal,colormap/bluered, horizontal style]
      \addplot[mesh,samples=150,domain=-2.5:2.5]
      {x};
      \addplot graphics
	[xmin=-20,xmax=20,ymin=-20,ymax=20]
   	{vely0500_pgf};
    \end{axis}
    \\
    };
  \end{tikzpicture}
  \caption{Velocity maps for \vsgr{} after five orbits.  The upper
    panel shows the $v_x$ component and the lower panel shows the $v_y$
    component.  The color scale is in units of $a\,\Omega$.
    }
  \label{fig:vel}
\end{figure}

\begin{figure}
  \centering
  \begin{tikzpicture}
    \begin{axis}[colorbar right,colormap/bluered, map style]
      \addplot[mesh,samples=150,domain=-16:-6] {x};
      \addplot graphics
	[xmin=-40,xmax=40,ymin=-40,ymax=40]
   	{binary_2d_5lev_hdf5_chk_0076_vect_pgf};
    \end{axis}
  \end{tikzpicture}
  \caption{
  Surface density and velocity distribution denoted by arrows for the
  same snapshot as shown in Fig.~\ref{fig:dens}.
  }
  \label{fig:vect}
\end{figure}
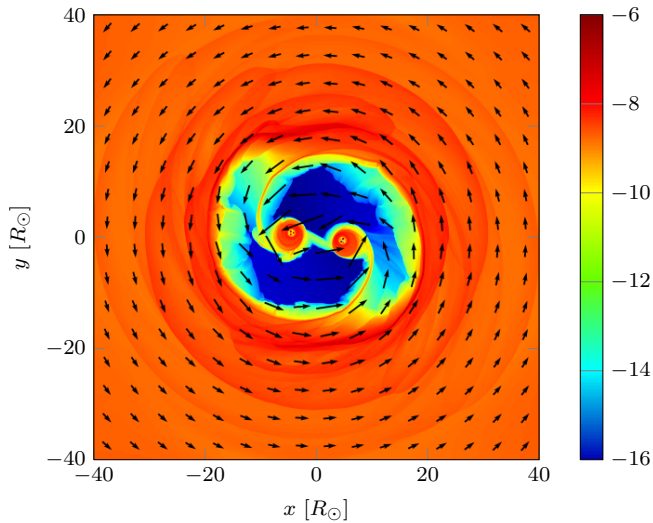

\begin{figure}
  \centering
  \begin{tikzpicture}
    \begin{axis}[colorbar right,colormap/bluered, map style]
      \addplot[mesh,samples=150,domain=-16:-6] {x};
      \addplot graphics
	[xmin=-20,xmax=20,ymin=-20,ymax=20]
   	{binary_2d_5lev_hdf5_chk_0076_stream_pgf};
    \end{axis}
  \end{tikzpicture}
  \caption{Surface density map in logarithmic scale with flow lines for
  the \vsgr{} simulation shown in Fig. \ref{fig:dens}.}
  \label{fig:stream}
\end{figure}
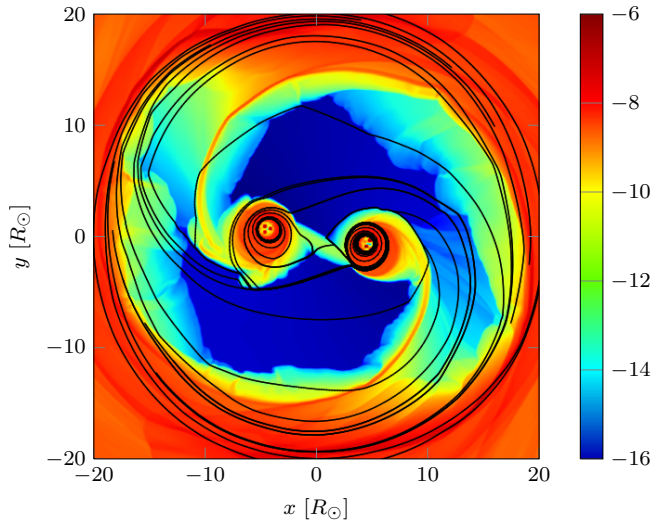

\begin{figure}
  \centering
  \begin{tikzpicture}
    \begin{axis}[colorbar right,colormap/bluered, map style]
      \addplot[mesh,samples=150,domain=-16:-6] {x};
      \addplot graphics [xmin=-40,xmax=40,ymin=-40,ymax=40]
   	{binary_2d_5lev_hdf5_chk_0076_block_pgf};
    \end{axis}
  \end{tikzpicture}
  \caption{Grid structure with four additional levels of refinement over
  the density map in logarithmic scale for the same snapshot as shown in
  Fig.~\ref{fig:dens}. The mesh is refined over the inner cavity and
  the spiral arms in the circumbinary disc.  The colour scale is the
  same as shown in Fig. \ref{fig:vect}.  }
  \label{fig:block}
\end{figure}
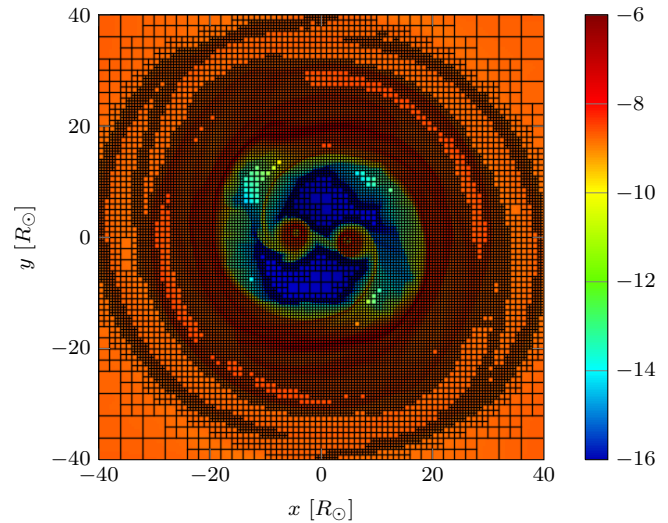

\begin{figure}
  \centering
  \begin{tikzpicture}
    \begin{semilogyaxis}[xlabel={$r\ [R_\odot]$}, ylabel={$\Sigma$},
      xmin=0,xmax=5,ymin=1e-2,ymax=1e3]
      \addplot[mark=none] table [x index=0,y index=1]
   	{figures/profile.txt};
      \addplot[dashed,mark=none] table [x index=0,y index=2]
   	{figures/profile.txt};
    \end{semilogyaxis}
  \end{tikzpicture}
  \caption{Averaged surface density in logarithmic scale as a function
  of radial distance in solar radii after 5 orbits.  The solid line
  corresponds to the averaged density around the primary and the dashed
  line is the density around the secondary.  There is a high density
  peak around each component since we do not consider accretion onto the
  stars.  }
  \label{fig:disc_profile}
\end{figure}
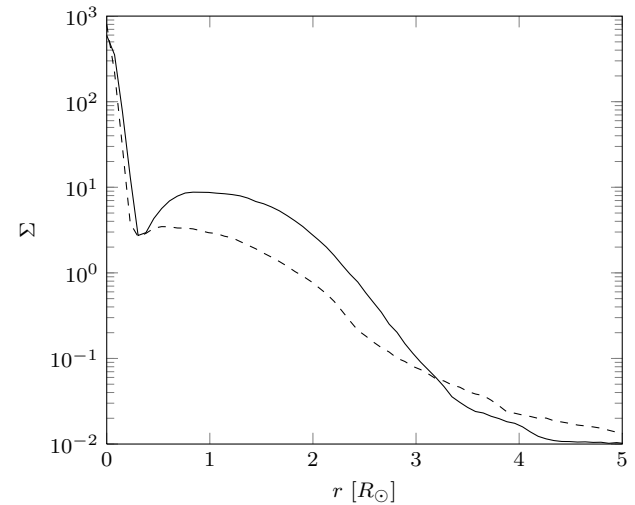

After about two orbits, we obtain a quasi-stationary and symmetric
configuration in the corotating frame of the system.
Figure~\ref{fig:dens} shows the surface density in logarithmic scale for
the \vsgr{} system after five orbits.  At this time, the surface density
within the inner hole is reduced about six orders of magnitude relative
to the density in the outer disc.  The corresponding velocity pattern is
shown in Fig.~\ref{fig:vel}.  Figure~\ref{fig:vect} shows the velocity
vectors in the inertial frame, and the corresponding flow lines in the
corotating frame are displayed in Fig.~\ref{fig:stream}. Although the
numerical viscosity in our code is small \citep{2007A&A...471.1043D},
diffusion into the inner hole can result from numerical viscosity or
shock formation.  Still, the tidal gap remains open for the duration of
our simulations although significant accretion continues onto the stars.

Two-armed spiral structures form in the circumstellar discs after
about one orbit, and a high-density arm joins both discs.  The
circumstellar discs are strongly perturbed by the accretion flows
through the inner cavity, although the spiral arms remain in a stable
configuration for the duration of our simulations.  The gravitational
softening affects the density distribution around the star but it does
not modify the accretion flows through the gap.

We have used a grid structure with four additional levels of refinement
shown in Fig.~\ref{fig:block}.  Almost the whole inner gap and stellar
components are refined and therefore have higher resolution.  The outer
regions of the disc have coarser resolution which allows us to speed up
the calculation.  From lower-resolution test runs involving more than 10
orbits we conclude that the pattern of gas flows and the position of the
edge of the circumbinary disc do not change significantly after about
five orbits.

\begin{figure}
  \centering
  \begin{tikzpicture}
    \begin{axis}[
      xlabel={$t$\ [d]}, ylabel={$M\ [10^{-9}\ M_\odot]$}
      ]
      \addplot[mark=none] table [x index=0,y index=1]
       	{figures/macc_00.txt};
      \addplot[mark=none,dashed] table[x index=0,y index=1]
       	{figures/V4046b.txt};
    \end{axis}
  \end{tikzpicture}
  \caption{Mass accreted around the primary (solid line)
  and secondary (dashed line) components
  as a function of time for the circular binary system \vsgr{}.
  }
  \label{fig:mdot}
\end{figure}
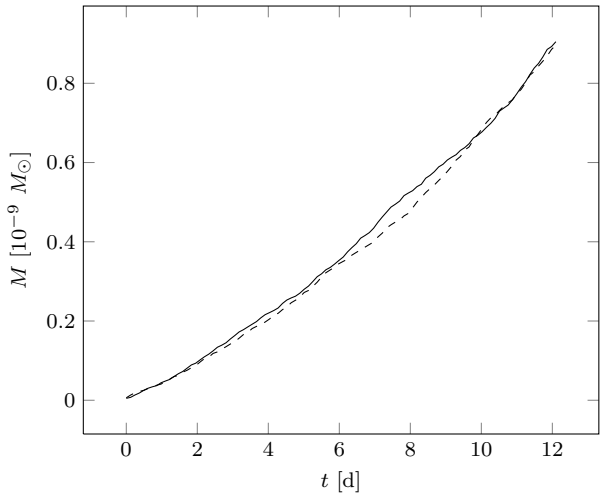

\begin{figure}
  \centering
  \begin{tikzpicture}
    \begin{axis}[colorbar right,colormap/bluered, map style]
      \addplot[mesh,samples=150,domain=-15:-6] {x};
      \addplot graphics
	[xmin=-20,xmax=20,ymin=-20,ymax=20] 
	{binary_2d_5lev_hdf5_chk_0053_pgf};
    \end{axis}
  \end{tikzpicture}
  \caption{Surface density map in logarithmic scale for a simulation of
  a circular binary system with a mass ratio of 1:5 and with the same
  notations as in Fig.~\ref{fig:dens}. }
  \label{fig:onetofive}
\end{figure}
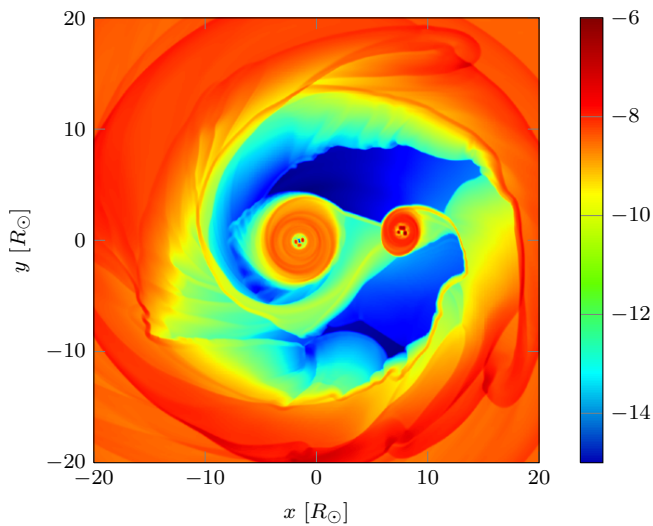

Figure~\ref{fig:disc_profile} shows the averaged surface density around
the stellar components as a function of radial distance. In this
simulation, mass is not accreted onto the stars, and the density peaks
in both discs reflect this accumulation of matter.  Since mass is not
removed from the disc, the profile does not correspond to a standard
accretion disc.  In addition, the density distribution in the inner disc
is influenced by the gravity softening.  
Therefore, the stellar cores
are excluded from the calculation of emission line profiles.
The circumstellar disc around the
primary is slightly more massive than the disc around the secondary.
If mass is accreted in the circumstellar discs, averaged accretion
rates onto the stellar cores are $\sim 2.8 \times 10^{-8} M_\odot\
\mathrm{yr}^{-1}$.  These values agree in order of magnitude with other
eccentric and circular binary simulations
\citep{2002A&A...387..550G,2004A&A...423..559G}. 
Fig.~\ref{fig:mdot} shows the accreted mass onto the primary against time
for the \vsgr{} model.

We then repeat our simulations of a circular binary but with components
of different mass. We set the mass ratio to 1:5 with the same primary
mass and mean distance between the components as in the \vsgr{} system.
As shown in Fig.~\ref{fig:onetofive} circular systems with low mass
companions also develop central gaps, non-axisymmetric mass accretion
and local circumstellar discs.  
The averaged mass accretion rate onto the primary is $3.1 \times 10^{-8}
M_\odot\ \mathrm{yr}^{-1}$, while the averaged mass accretion rate onto
the secondary is $2.2 \times 10^{-8} M_\odot\ \mathrm{yr}^{-1}$, and the
corresponding stellar disc is less massive. 

In the case of no mass transfer to the secondary, this component will
spiral inwards and eventually be swallowed by the primary.  It has
sometimes been speculated that mass accretion in systems with low mass
secondaries would occur preferentially to the secondary, and that this
process could prevent a rapid merging of the components. Our
calculations show that the overall process of accretion is the same as
in systems of mass ratio close to one. However, the gas flows to the
secondary will slow down the process of merging.  In the case of a
planetary mass companion one expects perturbations to
modify the orbital parameters of the planet that can lead to
large-amplitude eccentricity oscillations even for distant planets
\citep{1999AJ....117..621H}.

We conclude that binaries on circular orbits open gaps on short time
scales, and that substantial mass accretion takes place also in such
systems 
The accretion is non-axisymmetric, and gas flows from the
surrounding disc through the co-linear Lagrangian points, similar to
what was founr by \citet{2004A&A...421.1159S} from high-velocity
components in the Balmer emission lines of the object.  The accretion
rate is larger towards the primary, and its accretion disc is more
massive than the disc around the secondary in agreement with the
simulations by \citet{2005ApJ...623..922O}.

\subsection{Eccentric Binaries}

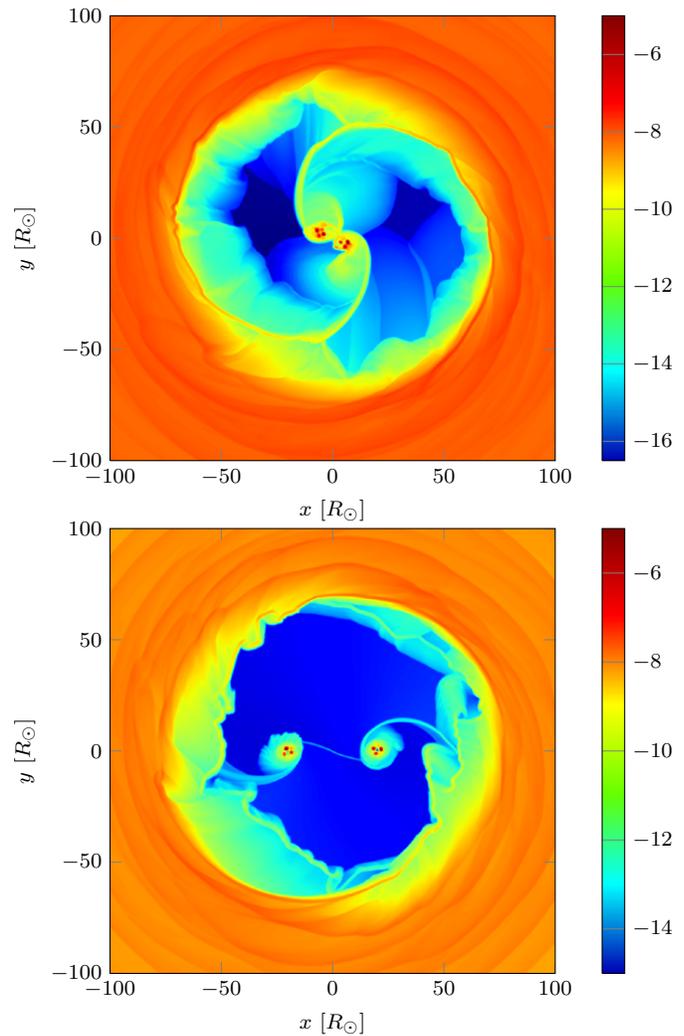
\begin{figure}
  \centering
  \begin{tikzpicture}
    \begin{axis}[colorbar right,colormap/bluered, map style,
      name=main plot]
      \addplot[mesh,samples=150,domain=-16.5:-5] {x};
      \addplot graphics
	[xmin=-100,xmax=100,ymin=-100,ymax=100] 
	{binary_2d_5lev_hdf5_chk_0097_pgf};
    \end{axis}
    \begin{axis}[at={(main plot.below south west)}, anchor=north west,
	colorbar right,colormap/bluered, map style]
      \addplot[mesh,samples=150,domain=-15:-5] {x};
      \addplot graphics
	[xmin=-100,xmax=100,ymin=-100,ymax=100] 
	{binary_2d_5lev_hdf5_chk_0075_pgf};
    \end{axis}
  \end{tikzpicture}
  \caption{Surface density maps in logarithmic scale for the eccentric
  binary system \dqtau{} in periastron (top) and apastron position
  (bottom). }
  \label{fig:RULupidens}
\end{figure}

In this section we present the results of high-resolution simulations of
a binary with components of nearly equal mass on a highly eccentric
orbit. We have selected the parameters for \dqtau{} as a case study with
the orbital parameters given in Table~\ref{table:DQTau}.  As in the case
of circular orbits we used a grid structure with a dynamically refined
grid with four additional levels of refinement.  The inner cavity is
refined and has higher resolution than the outer regions in the
circumbinary disc.
The surface density in logarithmic scale is shown in
Fig.~\ref{fig:RULupidens} at periastron and apastron.  The location of
the cavity rim in the system after it has reached a quasi-static
configuration is consistent with the prediction by
\cite{1996ApJ...467L..77A}.  The shape of the gap depends on the
eccentricity of the system and the disc viscosity
\citep{1991ApJ...370L..35A}.  The spiral arms in the outer circumbinary
disc become more pronounced than in the \vsgr{} high-resolution
simulations despite the presence of a wave damping region close to the
outer boundary, which are not seen in the circular binary simulations
with a mass ratio close to unity.

The size of the Roche lobe expands and contracts as the separation
between the components changes.  Hence, the circumstellar discs are
small compared with the circular binary simulations since the closest
approach between the stars is only 5.8 $R_\odot$. There is substantial
mass transfer between the discs as the secondary approaches periastron
and the spiral arms are truncated as shown in the early calculations by
\citet{1996ApJ...467L..77A}. Such periodic changes in the accretion
rates onto the stars have also been observed and may produce phase
modulations in the line emission \citep{1997ASPC..121..792R}.

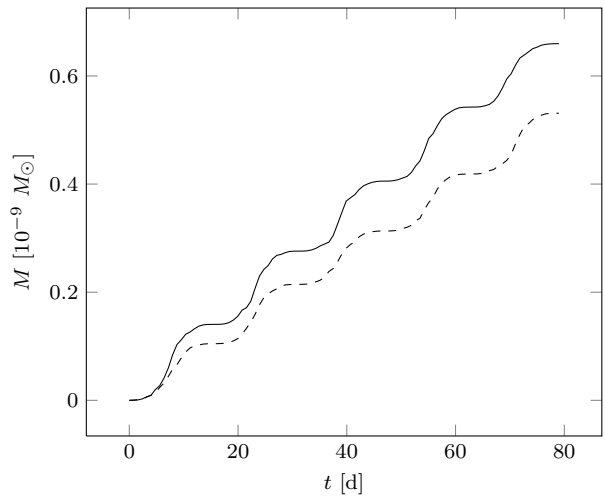
\begin{figure}
  \centering
  \begin{tikzpicture}
    \begin{axis}[
      xlabel={$t$\ [d]}, ylabel={$M\ [10^{-9}\ M_\odot]$}
      ]
      \addplot[mark=none] table[x index=0,y index=1]
       	{figures/macc_01.txt};
      \addplot[mark=none,dashed] table[x index=0,y index=1]
       	{figures/macc_02.txt};
    \end{axis}
  \end{tikzpicture}
  \caption{Time dependence of the accreted mass onto the primary 
  (solid line) and secondary (dashed line) components of the \dqtau{}
  system.
  }
  \label{fig:mdot-ecc}
\end{figure}

From our simulations we derive an averaged accretion rate onto the
primary of $8.1 \times 10^{-9}\ M_{\odot}\,\mathrm{yr}^{-1}$, and $6.2
\times 10^{-9}\ M_{\odot}\,\mathrm{yr}^{-1}$ for the secondary.
Averaged accretion rates onto the primary are higher than on the
secondary in agreement with the dual-grid eccentric binary simulations
by \citet{2002A&A...387..550G}.  However, the averaged accretion onto
the stellar components agree in order of magnitude with respect to
circular binary simulations presented in Section~\ref{sec:circular}.
Accreted mass onto the primary and secondary components
for the \dqtau{} model is shown in Fig.~\ref{fig:mdot-ecc}.
The accretion rates are larger during periastron passage
which can explain periodic changes in the observed luminosity. 
Averaged accretion rates in the orbital plane also agree within order of
magnitude with the values obtained from observed luminosities of
\dqtau{} by \citet{1998ApJ...492..323G}.

Orbital parameters of the binary system may be changed by stream flows
that modify the angular momentum of the orbit
\citep{1991ApJ...370L..35A}.  Although we evolve the binary orbit using
an N-body solver that includes the effects of the disc gravity, we are
not able to study the orbit evolution for sufficiently long time scales
in our model.  Considering the orbital evolution of the system would
require a simplified hydrodynamic description of the circumbinary disc
which is outside the scope of this study.

The patterns of gas flows around circular and eccentric binaries are
similar, but with some notable differences as seen when comparing
the density distributions shown in Fig.~\ref{fig:dens} and
Fig.~\ref{fig:RULupidens}. In the eccentric case shock fronts in the
streams from the disc edge to the stars produce narrow peaks in the
density map, and these can be traced as they spiral around each
stellar component before matter settle down in the local discs. These
discs are smaller than in the circular case since the close approach at
periastron prevents the formation of massive circumstellar discs.
In the eccentric case the central region is also connected with the disc
by bars of low density and perpendicular to the flow direction.
Only in the circular case with a mass ratio close to unity a stable
central bar connects the circumstellar discs.  Spiral patterns develop
in both the circular and eccentric cases, but are more pronounced in the
latter. In highly eccentric systems such as \dqtau{}, spiral arms are
formed around each stellar component, although the spiral patterns are
torn off as the system nears periastron.

\section{Comparisons with observations}
\label{sec:comparison}

High-resolution observations of \vsgr{} and \dqtau{} show that the
overall appearance of emission line spectra in close TTS binaries are
similar to those of single TTS. Both \vsgr{} and \dqtau{} show emission
features that are clearly connected to the stellar components, and which
vary in projected radial velocity as the stars move around the centre of
mass. This emission is certainly local, and the result of enhanced
chromospheric activity and accretion induced emission at the footprints
of the accretion funnels.  However, these stars differ from single stars
in that they show periodic fluctuations in  extended line wings of the
Balmer emission lines coming from warm gas flowing in the circumbinary
gap. This emission is a manifestation of tidal dissipation and viscous
heating in the accretion flow in the gap, and is powered also from other
energetic events in the regions and from photoionization from UV light
from the stars. The infrared excess emission comes from the cool, dusty
circumbinary disc, while there is no evidence of warm dust in the gap
from the energy distributions of the stars in question. 

In the case of \vsgr{}, \citet{2004A&A...421.1159S} found that the
Balmer lines have two broad emission components ($\Delta v \sim  170\
\kms$) related to gas moving in co-rotation close to the co-linear
Lagrangian points of the system and with projected velocity amplitudes
of $\sim 80\ \kms$. 
These components are optically thin, especially in the higher
Balmer lines,
since the Balmer decrement, as measured for the
higher Balmer lines, is more or less constant as a function of
velocity shift over the line wings.
The computations also indicate a much lower gas density in the
accretion flow than for accreting single stars, and since the flow has
gained a large spread in projected velocity, photons can easily escape
from any site. 

However, in our computations we did not consider details of heating or
cooling of the gas.  As a consequence the gas flows are cool over the
entire region, with temperatures below 1000 K, and no H emission can
arise. In order to make a first qualitative comparison between
calculated and observed line profiles for \vsgr{} we apply a different
equation-of state for the region inside the disc edge, where the H
emission is formed. Fig.~\ref{fig:dens2} shows the density and velocity structure
of such a modified model assuming a constant $T = 6000$ K in the gap.
Compared to the previous models the distributions of surface density and
velocity remain the same (see Fig.~\ref{fig:dens}).

\begin{DIFnomarkup}
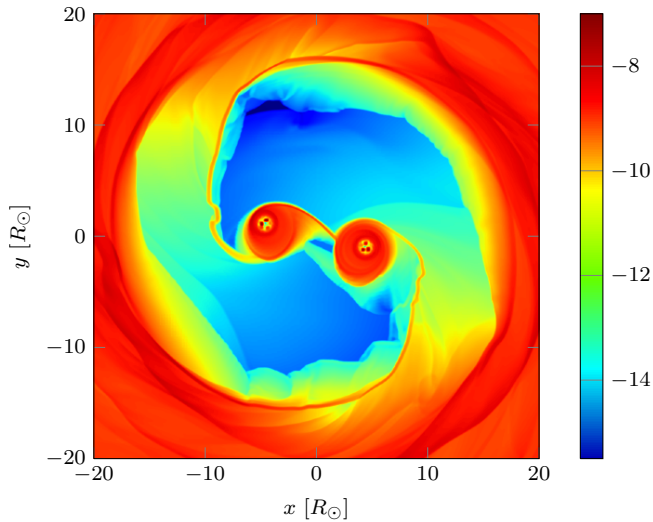
\begin{figure}
  \centering
  \begin{tikzpicture}
    \begin{axis}[colorbar right,colormap/bluered,map style]
      \addplot[mesh,samples=150,domain=-15.5:-7]
      {x};
      \addplot graphics [xmin=-20,xmax=20,ymin=-20,ymax=20]
   	{V4046_0076_pgf};
    \end{axis}
  \end{tikzpicture}
  \caption{Surface density map in logarithmic scale for a simulation of
  the system \vsgr{} after 5 orbits assuming a constant $T = 6000$ K in
  the gap.  }
  \label{fig:dens2}
\end{figure}
\end{DIFnomarkup}

In order to estimate expected emission line fluxes from the model we
assume that the disc has a fixed geometrical thickness.  The scale
height of the disc may change from the disc edge to the interior, but
these variations can be expected to be moderate and do not affect the
final qualitative comparison. Furthermore, the gas density is low and
the projected velocity range is large.  To characterize the line
emission as a function of orbital phase, we use the obtained
circumstellar density distribution and assume that the optically thin
emission proceeds from the accretion flows in the inner cavity. Since the
line emission can be assumed to be optically thin, in the higher Balmer
lines, except for the inner cores, we can assume that the line
emissivity per volume is proportional to the volume emission measure
($V\,n_e^2$). Hence, the total line flux within a certain velocity
interval is proportional to the surface density squared, $\Sigma^2$,
integrated over the area confined by the velocity interval in question.
No specific statements about the thickness of the accreting layer are
needed, since the total line intensities are scaled to match the model
profiles. 

\pgfplotsset{vproj style/.style=
    {enlargelimits=false,axis on top, width=.33\textwidth,
    height=0.819*.33*\textwidth,
    xlabel = {$x\ [R_\odot]$}, ylabel = {$y\ [R_\odot]$}}}

\pgfplotsset{profile style/.style=
    {enlargelimits=false,axis on top, width=.33\textwidth,
    xmin=-300,xmax=300, ymin=0, ymax=7,
    xlabel = {$v\ [\kms]$}, ylabel = {}}}

\begin{DIFnomarkup}
\end{DIFnomarkup}

\begin{DIFnomarkup}
\end{DIFnomarkup}

\begin{figure}
  \centering
  \begin{tikzpicture}
    \begin{axis}[xmin=0, xmax=1.,
      ylabel={$v_\mathrm{peak}$ [\kms]}, xlabel={Orbital Phase},
      xtick={0, 0.5, 1.}, xticklabels={0,$\pi$,$2\pi$},
      minor x tick num=3, minor y tick num=4]
      \addplot[mark=none]
      table[x index=0,y index=1] {peak_vel.txt};
      \addplot[mark=none]
      table[x index=0,y index=2] {peak_vel.txt};
      \addplot[mark=none,blue,dashed]
      table[x index=0,y index=3] {peak_vel.txt};
      \addplot[mark=none,blue,dashed]
      table[x index=0,y index=4] {peak_vel.txt};
    \end{axis}
  \end{tikzpicture}
  \caption{Peak velocities of the line profiles as a function of orbital
  phase for the \vsgr{} simulation. The solid lines show the 
  $v \Sigma^2$ weighted averages from the streams onto the primary and
  secondary components.  The peak velocities of the high velocity
  Gaussian components from the observations are shown by the dashed
  blue lines.
  Here, phase zero is defined defined as the time when the secondary
  passes the plane of the sky in the direction towards the observer.}
  \label{fig:vpeak}
\end{figure}
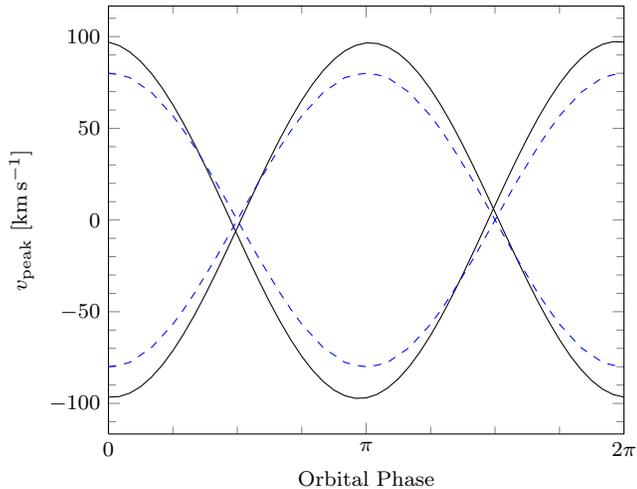

Our models predict that gas is moving through the co-linear Lagrangian
points of the system, just as inferred from the periodic changes
observed in the wings of the Balmer lines. We have computed theoretical
emission line profiles sampled over different areas inside the gap as a
function of phase, and as viewed in the line-of-sight to \vsgr{}. The
best agreement between observed and modelled line profiles, and their
variations, appears when in the model the emission is sampled only from
the area inside the disk edge, at an outer radius close to the 2:1
resonance at $1.6 a$, and extending inwards over the Lagrangian points
in question, to an inner radius of $0.8 a$. However, the exact locations
of these radii are not  very critical for obtaining a fair agreement
between observed and modelled peak velocity and line widths.
Fig.~\ref{fig:vpeak} compares the peak velocities as a function of phase as
derived from the model synthetic profiles with those observed by
\citet{2004A&A...421.1159S}, as described by their empirical
multi-component solution to the optically thin H$_8$, H$_9$ and H$_{10}$
Balmer lines, and excluding the stellar components.  The model amplitude
is 20\% larger than the peak velocity of the high velocity wings from
the Gaussian decomposition of the observed profiles.  However, the
relative phases of the peak velocities in the calculation agree well
with the observational data.

When also the central discs are included in the computations, strong
peaks in the line profiles at velocities extending to $200\ \kms$ are
created. These high velocity components are present at all phases, but
are not visible at all in the observed profiles. Hence, only by removing
the contributions from the stellar discs and their immediate
surroundings, we obtain synthetic line profiles that resemble the
observed ones for different phases.

We conclude that there is no observational evidence of any local discs,
but that a fair qualitative agreement between observed and modelled
properties can be obtained when these discs are removed. Then the
remarkable observed changes in the line profiles can be understood in
terms of perturbations in a circumbinary disc from which
non-axisymmetric gas flows are generated.  
The accreted material crosses the gap through the saddle points of the
potential down to a certain radius outside the orbiting discs.
The peaks of the observed line profiles at phases close to 0.0 and 0.5
in Fig.~\ref{fig:vpeak} differ from the model in the sense that
the model profiles are located at somewhat larger absolute velocities than
observed.

If the local discs are geometrically very thin their contribution to the
line emission might be very small and the corresponding high velocity
components might escape detection. 
As noted by \citet{1997AJ....114..781B}, most of the
spectral properties of TTS do not depend on whether the objects are
single or not. Narrow line components and continuous emission
(veiling) can be present in both cases and are usually related to
magnetospheric accretion.
We speculate that the absence of local
discs around the stars is related to magnetospheric accretion that could
dominate the accretion closer to the stars. Here, matter is channeled
along magnetic dipole fields to the polar regions of each star, and the
magnetic configuration co-rotates with the stars at relatively low
speed. 

Carlqvist and Gahm (unpublished) calculated the combined magnetic field
configurations resulting from two magnetic binary components. An example
is found in Fig.~23 in \citet{2001A&A...369..993P} for a binary with a
large mass ratio. However, calculations were made also for systems
similar to \vsgr{} and \dqtau{}, and with different assumptions on the
ratio of the magnetic moments. These calculations show that each star
maintains a pronounced local magnetosphere over an area in the disk
plane comparable to the size of the local disks present in our numerical
simulations. Further out from the stars the global magnetic field
strength declines roughly as r$^{-3}$, and is comparatively very weak at
the circumbinary disk edge. More detailed numerical simulations of the
combined magnetic fields in orbiting close TTS binaries, and their
interaction with surrounding plasma, have not been done so far.

Another aspect is that closer to the stellar surfaces the gas is
further heated and Balmer emission is not any more the dominant emission
component. The presence of hot central zones, in for example \vsgr{}, is
confirmed  by observations of variable FUV emission lines of
\mbox{He\,{\sc ii}}, \mbox{C\,{\sc iv}} and \mbox{Si\,{\sc iv}}
\citep{1986ESASP.263..107D} as well as significant X-ray emission
\citep{2006A&A...459L..29G}.

In the case of \dqtau{} our model indicates that there is substantial
accretion through the inner cavity onto the stars in agreement with
previous SPH simulations \citep{1996ApJ...467L..77A}.  Accretion onto
the stars increases significantly as they approach periastron in our
simulations of close eccentric systems. This can explain the periodic
outbursts in the emission lines and the continuous emission (veiling)
observed by \citet{1997AJ....114..781B}, and the sudden brightening at
millimetre wavelengths observed by \citet{2008A&A...492L..21S}.
\citet{1997AJ....114..781B} also discovered that H$_\alpha$ shows
enhanced red wings at phases near periastron.
Since no detailed line profile decomposition has been done so far, it is
unclear how these variations in line shape can be related to our model.
In addition, they found that mass ejection from the system could
occasionally occur in periastron. We note that
\cite{2005AJ....129..985H} found that the forbidden line emission does
not change with phase, indicating that this emission originates outside
the immediate vicinity of the stars. \citet{2001ApJ...551..454C} found
evidence that also cool molecular gas is moving inside the disc gap.

\section{Conclusions}
\label{sec:conclusions}

We have presented high-resolution simulations of mass accretion from a
circumbinary disc onto a close binary system.  We selected orbital
parameters from two T Tauri binaries, namely the circular spectroscopic
system \vsgr{} and the highly eccentric system \dqtau{}, in order to
compare model predictions with observed quantities. These systems have
components of nearly equal mass.  In addition we study a circular system
with a low mass companion.  The systems were evolved for long time
scales using an adaptive grid method and quasi-stationary solutions were
achieved in all cases after a few orbits.  We have considered the
dynamics and evolution of the discs and gas flows inside the
circumbinary gap.  The circumbinary disc edge is located close to the
2:1 resonance in agreement with the results from
\citet{1994ApJ...421..651A}.

These calculations confirm that significant gas accretion occur through
tidal gaps generated in such systems, in particular in the case of
circular orbits but with considerably lower surface densities compared
to the systems having highly eccentric orbits.  The gas flows are
non-axisymmetric in the studied systems, with two gas streams passing
the co-linear Lagrangian points on the way down to the stars, where
circumstellar discs develop.  However, the size of the gap stays
constant due to the equilibrium between viscous and resonant torques.
In the \vsgr{} case, spiral arms appear in the circumstellar disc
due to the strong gravitational torques from the close stellar
component, with a high density bar joining the circumstellar discs.  

We find that the local circumstellar disc around the primary is
significantly more massive than the disc of the secondary in our
simulations after several orbital periods when the system has reached a
quasi-static state.  Hence, mass accretion goes preferentially to the
primary, in accordance with the conclusions by \citet{2005ApJ...623..922O}.
In eccentric systems such as \dqtau{}, the circumstellar discs are cut
off as the system approaches periastron. Part of the material is then
accreted onto the stellar cores and accretion rates increase. These
periodic accretion changes and brightening events have been observed in
\dqtau{} \citep{1997AJ....114..781B} and confirmed by numerical
simulations \citep{2002A&A...387..550G}.  During apastron the stars
accrete mass from the circumstellar discs that form around each star.

We also used our simulation for a simple, qualitative comparison between
line emission from the region within the circumbinary disk and the
profiles observed by \citet{2004A&A...421.1159S}. We found that a fair
agreement with the velocity changes as a function of orbital phase
observed for the optically thin higher Balmer lines can be achieved if
an inner region surrounding the local circumstellar discs, is not taken
into account.
Our model can characterize the stream flows in the inner cavity of
circumbinary discs down to a certain radius from the centre. 
This shows that in T Tauri binary systems accretion may occur due to
gravitational interaction with the circumbinary disc. Since there is no
observational evidence for the presence of circumstellar discs, we speculate
that hydrodynamic accretion can be effective only down to a certain radius from
the centre, below which magnetospheric accretion is likely to take over the
accretion process. Our calculations of combined magnetic field configurations
around binary magnetic stars also show that each component is surrounded by
rather compact local magnetospheres, but that the field strengths decline
rapidly towards the circumbinary disk edge.

There are also a number of differences in observed and modelled line
profiles and the projected velocities at peak intensity, but only at
certain phases. The comparison shows that  gas emission is produced in a
larger region in co-rotation with the stars than predicted by the model.
Three-dimensional hydrodynamic models with self-consistent radiative
transfer will be necessary to better estimate line emission from this
object and provide constraints on the parameters of the system by
matching observed profiles.  Future calculations should also include a
more realistic mechanism to account for accretion onto the stars.

\section*{Acknowledgements}

MdVB was supported by a SAO predoctoral fellowship and a NOT/IAC
scholarship during the course of this project.  The code used in this
work is developed in part by the U.S.\ Department of Energy under Grant
No.\ B523820 to the Center for Astrophysical Thermonuclear Flashes at
the University of Chicago.  This work has made use of NASA's ADS
Bibliographic Services.  We thank Per Carlqvist for contributing with
calculations of combined magnetospheres, and Garrelt Mellema for comments and
suggestions.  We are grateful to the anonymous referee for helpful comments
that improved the manuscript.

\bibliographystyle{aa}
\bibliography{ads,preprints}

\label{lastpage}

\end{document}